# Interharmonic Power – A New Concept for Power System Oscillation Source Location

Wilsun Xu, *F.IEEE*, Jing Yong, *M.IEEE*, Horacio J. Marquez, *Senior M.IEEE,* Chun Li, *Senior M.IEEE*

*Abstract* — Power system oscillations are a significant concern for system operators, a problem that has grown due to the interconnection of inverter-based resources. To address this issue, various methods have been proposed to locate the sources of oscillations, which is essential for effective mitigation actions. A common characteristic of these methods is that they rely on phasor representation of oscillation phenomena. This paper takes a different approach by examining the actual voltage and current waveforms underlying the phasors. It is found that the presence of interharmonic components is both the necessary and sufficient condition for phasor oscillations. Moreover, the generation and propagation of interharmonic powers are identified as the true culprits behind power system oscillations and oscillatory instability. Based on these insights, two new methods are developed for locating oscillation sources: one for measurement-based monitoring applications and another for model-based system studies. These findings are validated through four field data-based and one simulation-based case studies.

## I. INTRODUCTION

Power system oscillation is always a concern to system operators. In the early 1980's, the phenomenon of inter-area low frequency oscillation triggered extensive research on power system stabilizers including their optimal locations [1,2]. In recent years, two developments have resulted in renewed interest in power system oscillation research: 1) the advance in measurement techniques such as synchrophasor-based wide area monitoring systems (WAMS); 2) the integration of inverter-based resources (IBRs) into power systems. Oscillations associated with wind and solar farms have emerged as a new concern to system operators [3].

Among the various research on power system oscillations, developing methods that can locate oscillation sources is of significant value, as finding sources is the prerequisite for mitigation actions [4]. There are two types of oscillation source location methods, measurement-based methods and model-based methods. The first type relies on the data collected by systems such as WAMS for source location and is a system monitoring application. The second type relies on computer models of a system to predict system stability and causes of instability. This is a system design application.

The research on measurement-based methods gained more attentions in recent years because of WAMS [4]. [5] provides an excellent review on the status of this line of research. Among the various methods, the dissipation energy-based methods seem to be most promising [6]. The method is based on heuristic formulation of synchronous generator instability. Thus, it is impossible to prove or disprove the method beyond case studies [7,8]. [5] concludes that more research is needed for measured-based source location methods.

The model-based methods were developed much earlier and a standard practice has been established. It is to perform eigen-analysis on the state equation of the linearized system model. The participation factors of unstable modes are then used to rank generators' participation in oscillations [2]. Since s-domain based methods are widely used to investigate the stability of IBR-connected grids [9,10], there are challenges to apply the method of [2] especially if the IBR models are obtained from lab tests. This situation has led to the use of a method developed for harmonic resonance analysis [11] for critical bus identification [12]. Although harmonic resonance and system instability may be related, they involve different mechanisms. The applicability of resonance methods is questionable. A more promising development is [13] which combines ideas of [11] and [14] and apply them to the s-domain $[Y(s)]$ matrix. Bus participation factors are proposed to identify locations that can excite instability more easily. However, this is not the same as finding oscillation source locations. Therefore, there is still a need to develop methods for model-based oscillation source location.

A common characteristic of the above research works (especially the measurement-based methods) is that they rely on the phasor representation of the oscillation phenomena. This paper approaches the problem by examining the actual voltage and current waveforms underlying the phasors. It is found that the existence of interharmonic components is the necessary and sufficient condition of phasor oscillation. Interharmonics are spectral components whose frequencies are not integer multiple of the fundamental frequency [15]. Furthermore, the generation and propagation of interharmonic powers are the true culprit behind power system oscillation in general and oscillatory instability in particular. Building upon these insights, two new methods for oscillation source location are developed, one for measurements and one for studies.

This paper is organized as follows: Section II introduces an alternative, interharmonic-based interpretation of synchronous generator (SG) oscillations, thus establishing an intuitive understanding on why interharmonics are the cause of SG oscillations. Section III presents mathematical proofs about the relationships of interharmonics, phasor oscillations and system stability. Based on these findings, Section IV

This work was supported by grants from the Natural Sciences and Engineering Research Council of Canada.
W. Xu, J. Yong and H.J. Marquez are with the University of Alberta, Canada (wxu@ualberta.ca, yongjingcq@gmail.com, hmarquez@ualberta.ca);
C. Li is with the Hydro One Limited, Canada (chester.li@hydroone.com)

introduces an interharmonic power-based oscillation source location method. Four case studies based on actual field data are used to validate the method. Section V introduces the concept of generator participation factor for model-based oscillation source location, along with case study results.

## II. A New Interpretation of Generator Oscillations

Synchronous generator (SG) oscillations are examined from the interharmonics perspective in this section. Three mechanisms of interharmonic generation are analyzed.

### A. Behavior of a SG Experiencing an Oscillating Voltage

Fig. 1a shows a 3Hz phasor oscillation. If we examine the underlying waveform instead, the oscillation is seen as a modulated 60Hz wave (Fig. 1b). By performing DFT on the waveform, we will notice there are two spectral components at 57Hz & 63Hz respectively (Fig. 1c). These are interharmonic components. It will be shown next that it is these two components that cause SG oscillations.

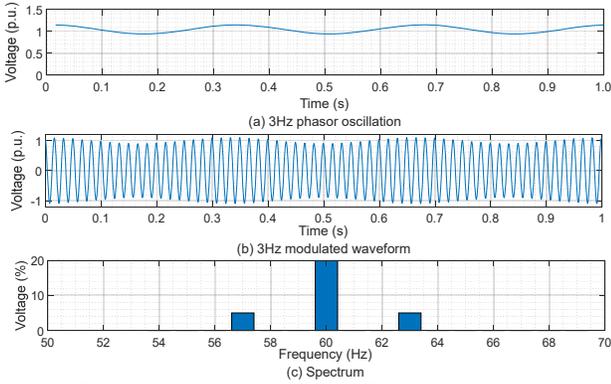

Fig.1: Oscillating V seen from phasor, waveform and spectral perspectives.

The oscillating voltage phasor can be expressed as follows:
$$\vec{V} = V(t)\angle\delta \quad V(t) = [1+m\cos(\omega_{os}t)]V_1$$

where $V_1$ is a constant representing the average magnitude of the phasor, $m$ is the magnitude of oscillation, and it is much smaller than 1. $\omega_{os}$ is the oscillating frequency. The actual time-domain voltage $v(t)$ underlying the phasor is

$$\begin{aligned}v(t) &= \sqrt{2}V_1[1+m\cos(\omega_{os}t)]\cos(\omega_1 t+\delta)\\ &= \sqrt{2}V_1\cos(\omega_1 t+\delta) + \sqrt{2}mV_1\cos(\omega_{os}t)\cos(\omega_1 t+\delta)\\ &= \sqrt{2}V_1\cos(\omega_1 t+\delta)\\ &\quad + \frac{mV_1}{\sqrt{2}}\{\cos[(\omega_1+\omega_{os})t+\delta]+\cos[(\omega_1-\omega_{os})t+\delta]\}\end{aligned} \quad (2.1)$$

where subscript "1" stands for fundamental frequency and $f_1$ denotes fundamental frequency throughout this paper. $\omega_1 = 2\pi f_1$. It can be seen that the actual voltage waveform is indeed composed of three waveforms with frequencies of $\omega_1$, $\omega_1+\omega_{os}$ and $\omega_1-\omega_{os}$ respectively. If we assume that the three phase voltages takes one of the following two forms:

$$\vec{V}_a = V(t)\angle\delta, \quad \vec{V}_b = V(t)\angle(\delta-120^o), \quad \vec{V}_c = V(t)\angle(\delta+120^o) \text{ or}$$
$$v_b(t) = v_a(t-T), \quad v_c(t) = v_a(t-2T), \quad T = 1/f_1$$

It can be shown that the resulting three-phase interharmonic components are in positive sequence if $f_{os} < f_1$. If a SG experiences these voltages, three rotating magnetic fields (m-fields) are produced in the airgap with rotation speeds of $N_s$, $(1+\omega_{os}/\omega_1)N_s$ and $(1-\omega_{os}/\omega_1)N_s$ respectively, where $N_s$ is the synchronous speed. This situation is shown in Fig. 2. Since the rotor speed is $N_s$, the m-field of speed $(1+\omega_{os}/\omega_1)N_s$ runs faster than the rotor. The SG behaves as an induction motor with respect to that interharmonic component. The m-field of speed $(1-\omega_{os}/\omega_1)N_s$ runs slower than the rotor. The *SG behaves as an induction generator* with respect to the $\omega_1-\omega_{os}$ interharmonic, and interharmonic power is generated and injected to the grid.

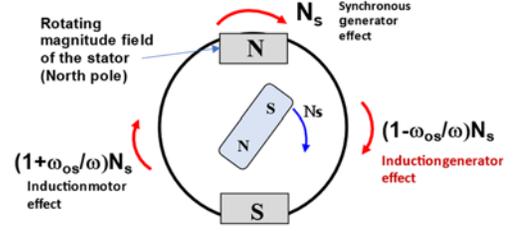

Fig. 2: Three rotating magnetic fields caused by an oscillating voltage.

The generation of interharmonic power at frequency $\omega_{IH}=\omega-\omega_{os}$ can be proven using the equivalent circuit of an induction motor shown in Fig.3. When a SG is superimposed with a voltage of frequency $\omega_{IH}$ while its rotor runs at $N_s$, the slip between the rotor and the stator $\omega_{IH}$ rotating fields is

$$s = \frac{N_{stator}-N_{rotor}}{N_{stator}} = \frac{(1-\omega_{os}/\omega_1)N_s - N_s}{(1-\omega_{os}/\omega_1)N_s} = \frac{-\omega_{os}}{(\omega_1-\omega_{os})} < 0$$

It can be seen that the equivalent rotor resistance $R_r/s$ is negative due to the negative slip $s$, thus interharmonic power is generated at $\omega_{IH}$ frequency by the SG.

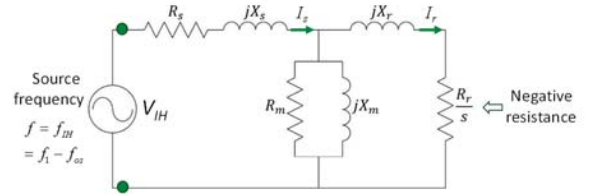

Fig.3: SG equivalent circuit at interharmonic frequency $\omega_{IH}=\omega-\omega_{os}$.

*If multiple SGs inject such interharmonic powers into the grid and the powers cannot be consumed by the grid, a sustained or growing oscillation will emerge in the power system. This is the mechanism of power system oscillation understood from the perspective of interharmonics.*

For large SGs, the magnetizing impedance can be neglected. The interharmonic power $P_{IH}$ can be estimated as

$$P_{IH} = R_s I_s^2 + \frac{R_r}{s}I_r^2 \approx (R_s + \frac{R_r}{s})I_s^2 \quad (2.2)$$

Eq.(2.2) reveals that a large current can lead to more interharmonic power generation. Since resonance can lead to a large current, resonance and oscillation are often related. One example is the sub-synchronous oscillations associated with series compensated lines. The finding also provides an intuitive explanation on why resistances in the grid can damp oscillations - since they consume interharmonic power.

## B. Interharmonic Generation by an Oscillating Field Current

This analysis is to prove that an (forced) oscillating field current produces interharmonic voltages at the SG terminal. The equivalent circuit of an SG stator is shown in Fig. 4 [16]. Here, $e$, $v$, and $i$ are the EMF (electromotive force), terminal voltage and current respectively. $L_{aa}$, $L_{ab}$ ... $L_{cc}$ are the self and mutual inductances of the three phase windings. The mutual coupling between the stator and field windings changes with the rotor position and needs to be expressed as

$$L_{af} = M_f \cos(\omega_1 t), \ L_{bf} = M_f \cos(\omega_1 t - \frac{2}{3}\pi), \ L_{cf} = M_f \cos(\omega_1 t + \frac{2}{3}\pi)$$

If the field current $I_f$ is forced to oscillate at $\omega_{os}$, it can be expressed as

$$i_f = I_f[1 + m\sin(\omega_{os}t)] = I_f + \Delta i_f(t)$$

The phase-$a$ EMF, $e_a$, in Fig. 4, can be derived as:

$$e_a = -\frac{d(L_{af}i_f)}{dt} = -M_f \frac{d}{dt}\{\cos(\omega_1 t) \times [I_f + \Delta i_f(t)]\}$$
$$= M_f I_f \{\omega_1 \sin(\omega_1 t)[1 + m\cos(\omega_{os}t)] - m\omega_{os}\sin(\omega_{os}t)\cos(\omega_1 t)\}$$
$$= M_f I_f \omega_1 \sin(\omega_1 t) + (M_f m\omega_1 - 0.5m\omega_{os})I_f \sin[(\omega_1 + \omega_{os})t]$$
$$+ (M_f m\omega_1 + 0.5m\omega_{os})I_f \sin[(\omega_1 - \omega_{os})t]$$

$e_b$ and $e_c$ have a similar form. It can be seen that the EMF contains two interharmonics of frequencies $\omega_1+\omega_{os}$ and $\omega_1-\omega_{os}$. These interharmonics will appear in current $i$ if the SG is connected to the grid. The above analysis proves that an oscillation DC field current actually produces interharmonic voltages at the SG terminal. Moreover, the magnitude of the fundamental frequency component $\omega_1$ does not oscillate.

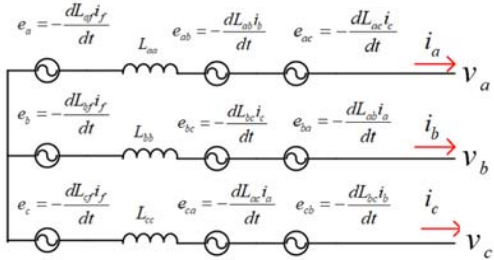

Fig.4: Equivalent circuit of a SG stator.

## C. Interharmonic Generation by an Oscillating Rotor

The single machine to infinity bus system has been widely used to explain SG oscillations as follows:

$$\vec{I}_{SG} = [E\angle\delta(t) - \vec{V}_\infty]/jX \quad \delta(t) = \delta_0[1 + m\cos(\omega_{os}t)] \quad (2.3)$$

where $E\angle\delta(t)$ is the phasor of the SG internal voltage. Its angle oscillates due to rotor oscillation. As a result, the SG current $I_{SG}$ also becomes an oscillating phasor, which in turn leads to oscillating power output from the SG.

Eq.(2.3) is actually an approximation of what really happens in a SG. As shown in Fig. 4, the true SG internal voltage produced by the field current is $e_a$. Since the rotor oscillates, the mutual coupling of stator and field becomes:

$$L_{af} = M_f \cos[\omega_1 t + \delta(t)], \quad \delta(t) = \delta_o[1 + m\cos(\omega_{os}t)]$$

Assuming a constant DC field current of $I_{f0}$ and considering $m$ is small, $e_a$ can be estimated as follows ($e_b$ and $e_c$ are similar):

$$e_a = -\frac{d(L_{af}i_f)}{dt} = -M_f I_{f0} \frac{d}{dt}\{\cos[\omega_1 t + \delta_0 + \delta_0 m\cos(\omega_{os}t)]\}$$
$$= M_f I_{f0} \sin[\omega_1 t + \delta_0 + \delta_0 m\cos(\omega_{os}t)][\omega_1 - m\delta_0\omega_{os}\sin(\omega_{os}t)]$$
$$\approx M_f I_{f0}\omega_1 \sin(\omega_1 t + \delta_0) + K_1 \sin[(\omega_1 + \omega_{os})t + \theta_1] + K_2 \sin[(\omega_1 - \omega_{os})t + \theta_2]$$

where $K_1$, $K_2$, $\theta_1$, and $\theta_2$ are constants. It can be seen that 1) the SG's internal voltage $e_a$ indeed contains interharmonics. 2) The $\omega_1$ component has a constant magnitude of $M_f I_{f0}\omega_1$ so it does not oscillate. Therefore, $\omega_1$ component cannot capture the characteristics of rotor oscillation strictly speaking. Eq.(2.3) is an approximation, and it worked in the past for SGs because of very low SG oscillation frequencies (<1Hz for inter-area oscillations). As IBRs are causing increased oscillation frequencies, it has become necessary to approach oscillation phenomena from their true cause, i.e. the interharmonics.

In summary, the analysis conducted on three distinct SG behaviors clearly demonstrates that interharmonics play a pivotal role in SG oscillations. Phasor oscillation is an appearance and, therefore, it has inherent limitations for investigating power system oscillations. As will be shown next, these findings are not confined to SGs; they also apply to IBRs and other types of generators.

## III. RELATIONSHIP BETWEEN PHASOR OSCILLATIONS AND INTERHARMONICS

To properly observe an oscillation event, data containing at least one full period of oscillation is required. Assuming the waveform is stationary (i.e. in a quasi-steady state) during this period, which is the case for sustained oscillations, this section will demonstrate that the presence of interharmonics is both a sufficient and necessary condition for phasor oscillation. Furthermore, the generation of interharmonic power is a necessary condition of small signal instability of a power system. These conclusions are independent of the types of generators involved.

### A. Sufficient Condition of Phasor Oscillation

The mechanism of an interharmonic causing phasor oscillation can be understood intuitively by examining a waveform that contains one interharmonic component. The waveform can be expressed as:

$$v(t) = v_1(t) + v_{ih}(t) = \sqrt{2}V_1 \sin(2\pi f_1 t) + m\sqrt{2}V_1 \sin(2\pi f_{IH}t + \varphi_{IH})$$

According to the PMU measurement standard [17], the phasor corresponding to the $k^{th}$ cycle of $v(t)$ is determined using the following formula:

$$\vec{V}_{phasor}(k) = \frac{1}{\sqrt{2}T_1}\int_{(k-1)T_1}^{kT_1} v(t)e^{-j2\pi f_1 t}dt \quad (3.1)$$

Fig. 5 shows the waveforms of $v_1(t)$, $v_{ih}(t)$, $v(t)$ and the magnitude of resulting phasor. It can be seen that the ±peaks of $v_{ih}(t)$ are not synchronized with those of $v_1(t)$. As a result, the two peaks add at the $k_1$-th cycle but subtract at the $k_2$-th cycle. Thus, the RMS value of $v(t)$ (with is proportional to the phasor magnitude) of $k_1$-th cycle is larger than that of $k_2$-th cycle, leading to the appearance of phasor oscillation.

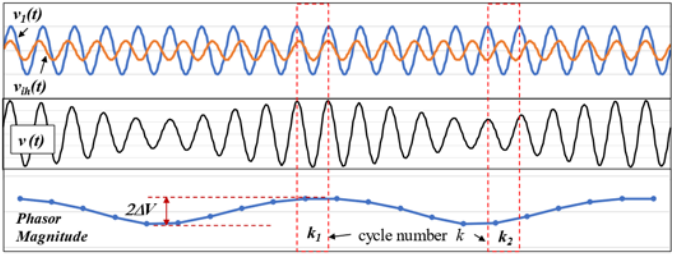

Fig. 5: Mechanism of an interharmonic causing phasor oscillation.

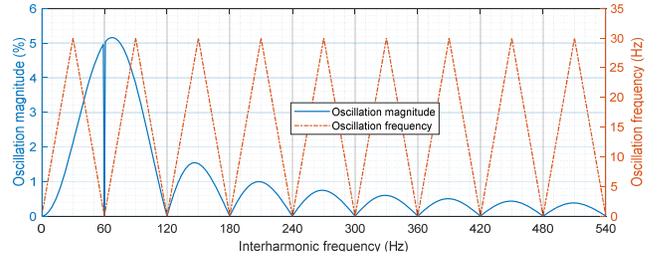

Fig. 6: Oscillation magnitude and frequency of phasor as affected by $f_{IH}$.

The phasor magnitude can be determined analytically as

$$|\vec{V}_{phasor}(k)| = V_{phasor}(k) = \sqrt{A^2 + B^2}$$

$$A = V_1\{1 + \frac{2mf_1 f_{IH}}{\pi(f^2_1 - f^2_{IH})}\sin(\frac{f_{IH}}{f_1}\pi)\cos[\frac{f_{IH}}{f_1}\pi(k-1) + \varphi_{IH}]\}$$

$$B = V_1\{\frac{2mf_1 f_1}{\pi(f^2_1 - f^2_{IH})}\sin(\frac{f_{IH}}{f_1}\pi)\sin[\frac{f_{IH}}{f_1}\pi(k-1) + \varphi_{IH}]\}$$

Since $m$ is much less than 1, the term of $m^2$ in above equation can be omitted. Thus,

$$V_{phasor}(k) \approx V_1\{1 + \frac{4mf_1 f_{IH}}{\pi(f^2_1 - f^2_{IH})}\sin(\frac{f_{IH}}{f_1}\pi)\cos[2\pi f_{IH} t + \varphi_{IH} - \frac{f_{IH}}{f_1}\pi]\}^{1/2}$$

$$\approx V_1\{1 + \frac{2mf_1 f_{IH}}{\pi(f^2_1 - f^2_{IH})}\sin(\frac{f_{IH}}{f_1}\pi)\cos[2\pi f_{IH} t + \varphi_{IH} - \frac{f_{IH}}{f_1}\pi]\}$$

$$= V_1\{1 + \Delta V \cos[2\pi f_{IH} t + \varphi_{IH} - \frac{f_{IH}}{f_1}\pi]\}$$

The above confirms that the phasor magnitude oscillates, and the degree of oscillation is $\Delta V$. $\Delta V$ can be estimated as:

$$\frac{\Delta V}{V_1} \approx \left|\frac{2m}{\pi}\frac{f_1 f_{IH}}{f^2_1 - f^2_{IH}}\sin(\frac{f_{IH}}{f_1}\pi)\right| \quad (3.2)$$

The frequency of phasor oscillation can be estimated as

$$f_{os} = |f_{IH} - hf_1| \quad (3.3)$$

where $h$ is the harmonic number nearest to $f_{IH}$. Assuming $m=5\%$, the degree of phasor oscillation $\Delta V/V_1$ as a function of $f_{IH}$ is shown in Fig. 6. The main findings are:

- The phasor magnitude has become oscillatory due to $f_{IH}$ component, *even though both $f_1$ and $f_{IH}$ components don't oscillate*. In other words, an oscillating phasor does not necessarily imply an oscillating $f_1$ component. Case studies later will further show the importance of this finding,
- The phasor oscillation frequency $f_{os}$ is in the range of 0 to $f_1/2$ regardless of $f_{IH}$ values. For example, $f_{IH}$=36Hz and 84Hz yield the same $f_{os}$ of 24Hz. This is related to the Nyquist sampling theorem. It implies that $f_{os}$ cannot be used to estimate $f_{IH}$,
- There is no phasor oscillation if $f_{IH}$ is equal to harmonic frequencies. In other words, only interharmonics cause phasor oscillation.

The case of two interharmonic components has been investigated in [18]. It is reasonable to expect that phasor oscillation is more likely to occur if there are more interharmonic components. But the oscillation may not take the single-tone sinusoidal form (see case study #4).

### B. Necessary Condition of Phasor Oscillation

The existence of interharmonics is also a necessary condition of phasor oscillation. This section will use a logical reasoning process to justify the assertion. (A formal mathematical proof is available but is omitted due to space limitation). Without loss of generality, a 60Hz voltage whose phasor oscillates at, for example, $f_{os}$=5Hz is considered. A basic assumption of Fourier transform is that the data used for the transform should cover at least one period of a signal. In this case, the period is $T_{os}=1/f_{os}=0.2sec$, so the width of data window is $T_{os}$. The result of transform can be expressed as:

$$v(t) = v_{DC}(t) + v_{w1}(t) + ... + v_{wk}(t)... \quad k = 1, 2, 3...$$

Subscript "$w$" emphasizes that the result is derived using $T_{os}$=0.2sec window. $V_{wk}(t)$ is called a spectral component since its frequency is $kf_{os}$ which is not necessarily equal to a 60Hz harmonic frequency. For this Fourier series, the following reasoning process can be applied:

1) Each spectral component $v_{wk}(t)$ represents a constant magnitude sinusoidal wave per the definition of Fourier series. In other words, each component does not have an oscillating phasor characteristic,
2) When these components add together to form $v(t)$, the only way to get an oscillating phasor of $v(t)$ is that the ±peaks of some of the components are not synchronized with the 60Hz wave, as illustrated in Fig. X,
3) Since only interharmonics have their ±peaks not synchronized with the 60Hz wave, one can conclude that the existence of at least one interharmonic component is the necessary condition of phasor oscillation.

Another way to understand the above logic is as follows: the fundamental frequency component cannot possibly oscillate because, if it did, it would have to contain frequencies other than $f_1$ and thus it is not a fundamental frequency component. Since harmonics cannot cause phasor oscillation either (per Section III.A), interharmonics are the only components capable of causing phasor oscillation.

### C. Necessary Condition of System Instability

For small signal stability analysis, generators which include SGs and IBRs can be modeled as an impedance in parallel with a current source in s-domain. The impedance is obtained by linearizing its dynamic equations around an operating point. The current source ($\Delta i_{gen}$) represents perturbations experienced by the generator. The grid can also be modeled similarly. The interconnection of the generator to the grid in the form of s-domain impedance model is shown in Fig. 7a.

The system of Fig. 7a can be rearranged into a standard feedback control structure consisting of transfer functions $H_1(s)$ and $H_2(s)$ (Fig. 7b). The stability characteristics of such a system have been well studied. For example, [19] has proven the following (including multiple $H_1$ subsystems):

*If both systems $H_1(s)$ and $H_2(s)$ are passive and one of them is (i) strictly passive and (ii) has finite gain, then output v and i are bounded whenever $u_1$ and $u_2$ are bounded.*

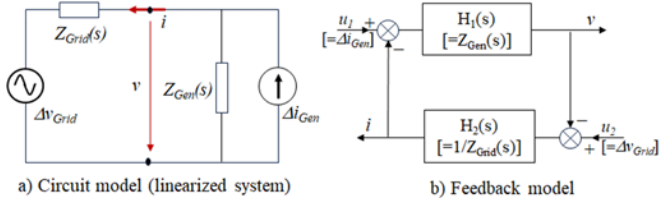

Fig. 7: Feedback model of a generator-grid interconnection.

A passive system $H(s)$ means its frequency response $H(j\omega)$ plotted in the form of magnitude vs phase is in the right half plane. For a circuit, it means there is no negative resistance or conductance at any frequency. Strictly passive means the energy dissipation in the system is more than energy input into the system. Finite gains are satisfied in circuits with resistances even under resonance condition, i.e. the peak of resonance is limited. This stability theorem essentially asserts that the existence of negative resistance or conductance at some frequencies is the necessary condition of instability.

In fact, the existence of negative incremental resistance behavior in IBRs caused by inductor generator effect or PLL block has been recognized as the cause of IBR oscillatory instability [3,20]. Section II.A has shown that the SG oscillation is also caused by negative resistances.

Since negative resistances can produce powers at frequencies where they are negative and the frequencies are very likely to be interharmonic frequencies, we can conclude that the generation of interharmonic power is the necessary condition of small signal instability. The instability usually manifests as growing phasor oscillation. If the interharmonic energy in the system does not either grow or dissipate, the oscillation becomes a sustained oscillation.

### D. Natural Oscillation versus Forced Oscillation

There are mainly two types of oscillation phenomena in power systems [4]. One type is natural oscillations which are caused by interactions of various components in the system. The other type is called forced oscillations which are due to external forces. The relationship between interharmonics and phasor oscillation revealed in Section III.A&B are applicable to both types of oscillations.

The finding of Section III.C is applicable to natural oscillation. It shows that the generation of interharmonic power inside the system via the mechanism of negative resistance is the necessary condition of such oscillations. The case of II.A is an example. Since the interactions of various components and controllers are involved in interharmonic generation, there is no single oscillation source strictly speaking. The sources are called participants according to [2].

The forced oscillation involves an external force. Section II.B shows an example where oscillating field current is the force. These external forces are equivalent to interharmonic current or voltage sources imposed on the system (See Sections II.B&C, Case #4 and [8]). Since there is typically a single source involved, it is reasonable to call the component involving the external force the oscillation source. Clearly, the interharmonic current or voltage sources need to produce powers to drive the flow of interharmonics in the system.

There could be rare occasions where the external force excites the natural oscillation of a system. Interharmonic power may be produced by both the equivalent interharmonic current or voltage source as well as the negative resistances.

## IV. A MEASUREMENT-BASED METHOD FOR SOURCE LOCATION

The findings of Sections II and III show that it is the interharmonic powers that drive system oscillations. Thus, the amount of active interharmonic powers generated by various candidate sources can be used for source location. The method and its application cases are presented in this section.

### A. Algorithm of Oscillation Source Location

The proposed oscillation source locator uses the voltage and current waveform data collected at the interface points of generating plants and grid. Various monitors are installed nowadays to collect such data including synchronized waveform data [21]. The simplest version of the data algorithm is as follows:

1) Perform DFT on the waveform data by using a moving window and extract significant interharmonic components from the results. [22] provides information on practical signal processing issues involving interharmonics,
2) Calculate three-phase active powers for the interharmonics of interest, one data window by one data window,
3) If the powers oscillate, compute the average powers or the accumulated powers (i.e. the energies) over the period of interest. Both can measure the net effect of interharmonic power generation/consumption over a period,
4) Compare the interharmonic powers and/or energies of all monitored generators obtained for the period of interest. Based on the amount of power and/or energy generated determine if a generator is a source/participant and its degree of contribution to the oscillation event.

The above algorithm can be implemented in distributed online monitoring systems easily. It can also be made into an offline tool for troubleshooting and postmortem investigations. Individual generators in a power plant can also be monitored to determine if a specific generator is the source. Data synchronization is required if interharmonic powers of multiple sites are to be compared [21] (also see Case #2).

### B. Field Measurements Based Case Studies

The thesis of interharmonic causing phasor oscillation and the method of oscillation source location are verified using four sets of field data collected from various oscillation sources. Field data supports more rigorous verification as the method is tested with realistic conditions. Due to space limitation, we can only present the key results.

Case 1: Synchronous Generator

This case is about a 60MW, 14kV SG. The SG experiences oscillation if its voltage goes below 0.96pu due to voltage control issue, so the SG is the source of oscillation. The oscillation can be seen from the waveforms measured at the SG terminal (Fig. 8). Visual estimation indicates that the oscillation frequency is about 1.8Hz.

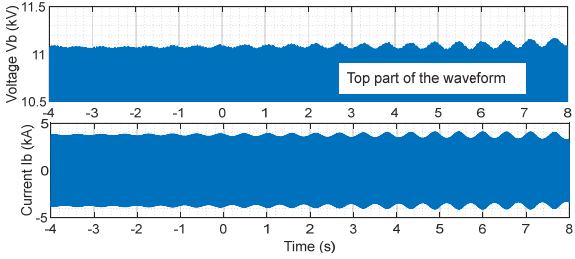
Fig. 8: Field measured voltage and current waveforms of a SG.

Fig. 9 shows the spectrum of a 4-second data window. Two dominant interharmonic components, around 58.2Hz and 61.8Hz, can be noticed. They are more easily observed from the current spectrum as voltages are stiffer than currents. Both interharmonics are in positive sequence and the frequencies agree with the formula $f_{IH}=f_1 \pm f_{os}$. The active power shown in the bottom chart confirms that interharmonic power is indeed generated around 58.2Hz as predicted in Section II.A.

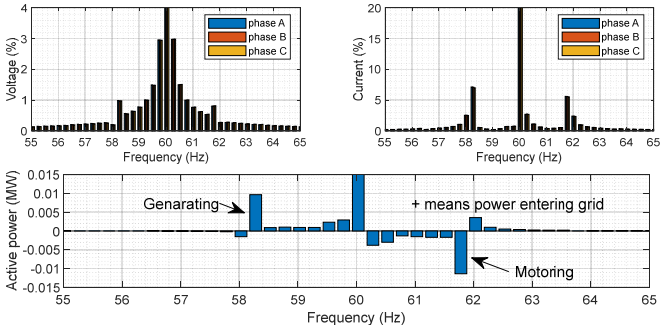
Fig. 9: Spectrums of voltage, current and power of the SG.

The SG's field voltage and current are also analyzed. Fig. 10 shows the field voltage. One can notice that there is a major interharmonic around 1.8Hz and it is even higher than the DC component. This situation signals that the field excitor's controller was indeed not working properly and should be the cause of oscillation. The figure also confirms the analysis of Section II.B.

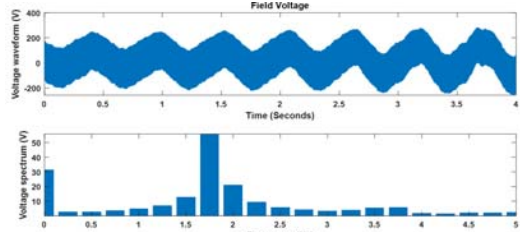
Fig. 10: Waveform and spectrum of the SG's field voltage.

Fig. 11 reveals more useful information. It shows the trend of (three-phase) phasor power, 60Hz power, total power and interharmonic powers. Phasor power is the one measured by SCADA or PMU [17]. Total power is the true active power flowing out of the generator per the law of physics:

$$P_{total} = P_1 + P_2 + ... + P_k + ...$$

where $P_k$ is the active power related to $k^{th}$ spectral component and $P_1$ is the same as the 60Hz power. According to prevailing understanding (including these authors), the phasor power is supposed to represent 60Hz power. However, the figure shows that the true 60Hz power (and even the total power) exhibits little oscillation, while the phasor power has about 10% oscillation at the last second of the plot.

The significance of this finding is the following: power oscillation does not occur at fundamental frequency. Phasor power oscillation is the byproduct of interharmonics. Even a small amount of interharmonics can modulate the voltage and current (see Section III.A), causing large apparent oscillations of the power calculated from them. But in reality, the 60Hz power is almost constant. In addition, the finding provides an alternative explanation on why the dissipation energy-based method [6] is not reliable. The method is based on phasor power. Since phasor power combines all spectral components and the effect of aliasing (see Section III.A), it is impossible to isolate the true culprit, one or two specific interharmonics, for proper source location.

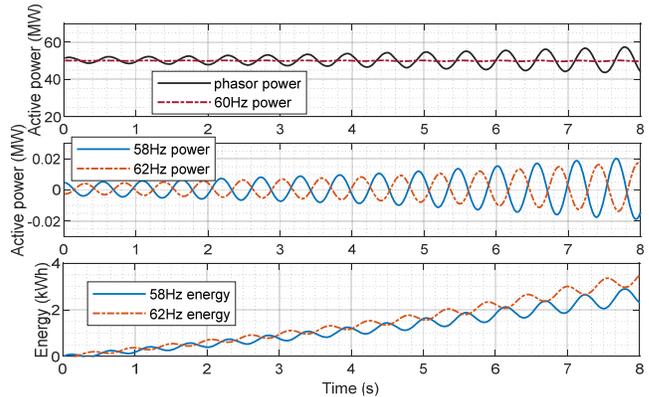
Fig. 11: Trend of phasor, 60Hz, total and IH powers, and IH energy.

The second finding is that powers are generated at both 58.2Hz and 61.8Hz and they oscillate, which appears contradicting to the prediction of Section II.A. This is caused by the following: the rotator speed is assumed to be a constant in Section II.A to facilitate explanation. In reality, the rotor also oscillates so both interharmonics can generate power and their powers can oscillate. In addition, the oscillating field current also causes interharmonic power generation (Section II.B). This is another valuable finding made possible with field data. The net power is positive, so the SG is the source. Since the interharmonic power oscillates in this case, its energy (Fig. 11c) or average power can be a more distinct indicator.

Case 2: Wind Farms

This case is about multiple IBRs connected to a transmission grid (Fig. 12). The opening of breaker B triggers an unstable oscillation in the area. Adequate data is available only for wind WF2 and WF3. The goal here is to determine which wind farm contributes more to the oscillation.

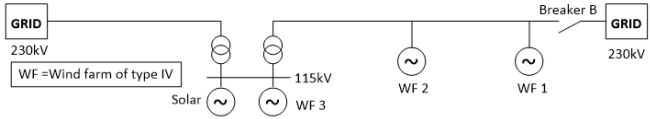
Fig. 12: A real-life grid with multiple IBRs.

Fig. 13 shows the current waveforms of both farms. Note that only 5 seconds of WF3 data is available. The data of the two sites are aligned using the internal clock of the monitors. The spectrum of WF2 and WF3 are shown in Fig. 14. It can be seen that there are two dominant interharmonics and they have the same frequencies at both sites. The frequencies agree with formula $f_{IH}=f_1 \pm f_{os}$. The interharmonics are positive sequence.

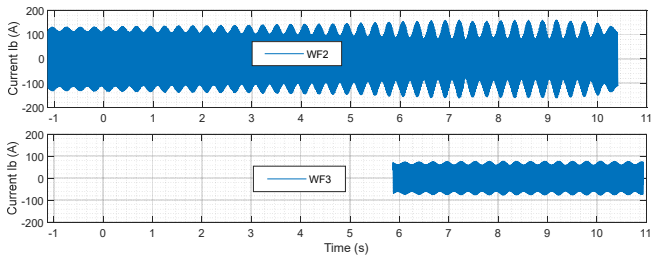
Fig. 13: Current waveforms of two wind farms (WFs).

The trends of various powers are shown in Fig. 15 for both farms. The results again show that phasor power oscillates and the 60Hz power does not. (Total powers are not shown as they overlap with the 60Hz powers). Both interharmonics generate power at both sites, so both farms are sources. However, WF2 generates more power (in % and kW) than WF3 so WF2 is the main contributor to the oscillation among the two.

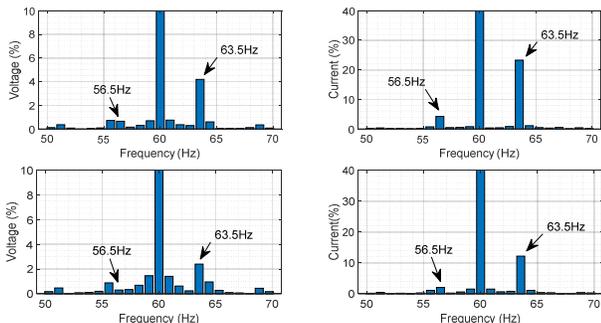
Fig. 14: Voltage and current spectrums of WF2 (top) and WF3 (bottom).

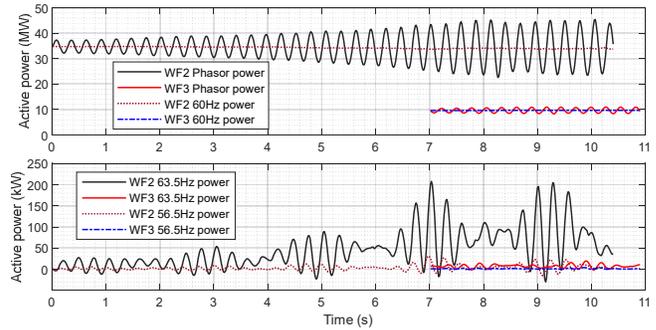
Fig. 15: Trend of phasor, 60Hz and interharmonic powers of two WFs.

Fig. 16 compares the phasor, 60Hz and interharmonic currents of WF2. It can be seen that the 60Hz current does not oscillate but the phasor current does. In addition, the phasor oscillation grows as the interharmonic currents increase, which gain confirms interharmonics modulation is the cause of phasor oscillation.

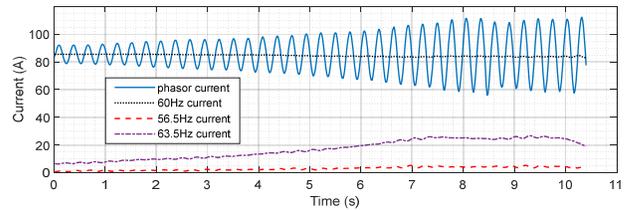
Fig. 16: Trend of phasor, 60Hz and interharmonic currents of WF2.

Case 3: Solar Farm

This case is about a 10MW solar farm connected to a 44kV feeder. The energization of a shunt capacitor nearby triggered oscillation of the solar farm's current as shown in Fig. 17. Visual estimation of the waveform gives an oscillation frequency of ~23Hz. The current spectrum and interharmonic powers are shown in Fig.18. Since there are 10.5 cycles of data available, the spectral resolution is only 6Hz (=60/10). Such a low resolution makes it difficult to determine the interharmonic frequencies precisely. In spite of this challenge, the results show that there is a large interharmonic around 84Hz and its power enters the grid, meaning the solar farm is a source of oscillation. Inverter PLL malfunction was suspected as the cause [23]. According to Eq. (3.3), an 84Hz interharmonic frequency corresponds to an 24Hz oscillation frequency which is very close to the visual estimation result.

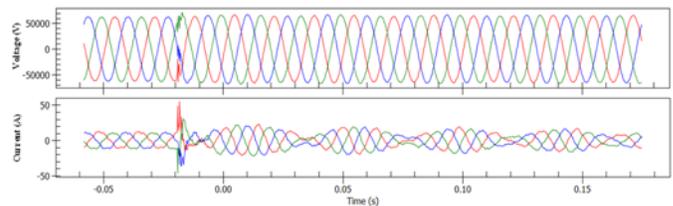
Fig. 17: Three-phase voltage and current waveforms of a solar farm.

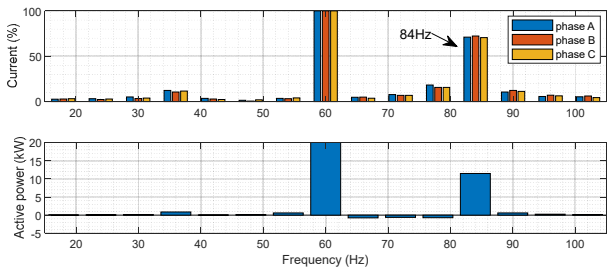
Fig. 18: Spectrums of current and interharmonic power of the solar farm.

Since this case involves fully rated inverters, the voltage and current seen by the inverter controller, i.e. the dq-transformed *v* and *i* are also examined. According to common understanding, an oscillating abc phasor will become an oscillating DC waveform in dq frame. The dq current shown in Fig. 19 reveals that the DC oscillation is actually caused by the interharmonic components around 24Hz. This frequency agrees with the theoretical analysis of the dq transform.

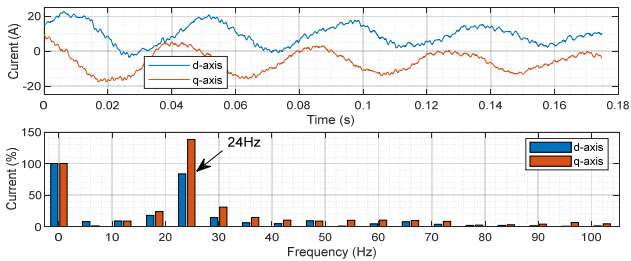
Fig. 19: Waveforms and spectrums of dq currents.

Case 4: Plant with Variable Frequency Drives (VFDs)

This case involves a 4.16kV pumping station and its VFDs shown in Fig. 20. Nearby customers such as C complained of light flickers for years. Locating the flicker source was not successful until multi-cycle waveform data was captured and its spectrum analyzed. Fig. 21 shows sample results when the VFD operated at 30Hz. We can see that the voltage exhibits a multi-tone oscillation. There are two dominant interharmonics at 117Hz and 237Hz respectively.

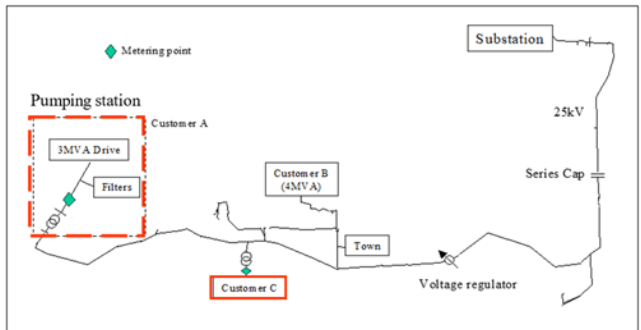
Fig. 20: Network diagram of Case 4.

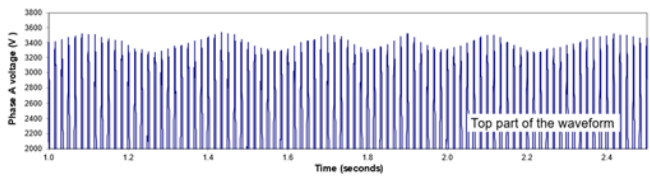

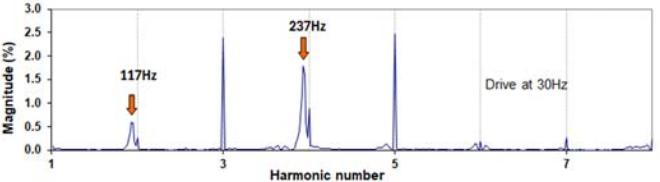
Fig. 21: Waveform and spectrum of the voltage at the pumping station.

Fig. 22 shows the measured 60Hz and interharmonic powers averaged over a 4 second measurement duration. The 60Hz and interharmonic powers have opposite signs, which means interharmonic powers are injected into the grid. Therefore, the plant is the source of oscillation. This conclusion is also supported by the work of [14] which shows a VFD can produce interharmonic currents by its nature of operation. Further investigation revealed that the impedances of the grid and plant filter formed a parallel resonance circuit. The 237Hz current source of the VFD excited the resonance and caused excessive voltage oscillation. The problem was solved by re-configuring the plant filter. This is a forced oscillation case since the source is external to the utility grid.

Fig. 22: Interharmonic power and its direction.

*C. Summary and Remarks*

Multiple real-life case studies in this section have confirmed the usefulness of the interharmonic power-based oscillation source location method, for both natural and forced oscillations. New information extracted from the field data further supports the thesis that phasor oscillations are caused by interharmonics.

It is worthwhile to note that there are two perspectives of system oscillations. For utility operators, it is a stability concern that affects grid operation. For utility customers, it is a power quality concern called voltage fluctuation or flicker. Interharmonics have been a known source of voltage flicker for some time [25]. The work on voltage flicker and interharmonics not only lends more support to the thesis of this paper but also provides useful experiences on the signal processing issues involving interharmonics [22,24-28].

V. A MODEL-BASED METHOD FOR SOURCE LOCATION

The objective of model-based methods is to determine the participants of natural oscillations. There is no need to do source location for forced oscillations since the source is to be modelled for study and thus it is already known.

*A. Determination of Poles Using 2D Frequency Scan*

For small-signal stability studies, representing IBR units using s-domain impedance model $Z_{Gen}(s)$ have gained wide acceptance [9,10]. It has led to s-domain, nodal matrix-based stability analysis of IBR-connected grids such as [13], i.e. investigating the stability characteristics of the system below:

$$[Y(s)][\Delta V] = [\Delta I] \quad (4.1)$$

where $[Y(s)]$ is the s-domain nodal matrix of the grid including IBR units. Symbol $\Delta$ is to emphasize this is a linearized model, and $[\Delta I]$ represents perturbations. Since interharmonics are closely related to instability and oscillation, the above model must cover interharmonic frequencies. Thus, it is an EMT-based, not phasor-based model. For example, the series impedance of a line must be modelled as $R+sL$, not as a constant $R+j\omega_l L$. Fortunately, [14] has established a foundation for EMT-based [Y(s)] models. It further shows that the poles of the system can be found by solving:

$$\det[Y(s)]=0 \quad (4.2)$$

i.e. the poles are the s values that make at least one of the eigenvalues of $[Y(s)]$ equal to 0. If one of the poles are on the right-hand side of the s-plane, the system is unstable.

[14,13] propose to use an iterative search method to solve Eq.(4.2). The method faces at least two challenges: 1) uncertainty of convergence characteristics and 2) difficulty to find all unstable poles. Inspired by the frequency-scan technique widely used for harmonic resonance analysis, this paper proposes to perform a simple 2D frequency scan to find the poles instead. The result can also reveal the overall stability profile of the grid in the form of a surface or contour (see Fig. 23&25). The scan is performed as follows:

1) Define the scan range as $\alpha=0:\alpha_{max}$, $f=f_{min}:f_{max}$, which is the right-hand side of the s-plane from $f_{min}$ to $f_{max}$,
2) For each pair of $\alpha$ and f value, let $s=\alpha+j2\pi f$,
3) Calculate the impedances of all network components including IBR units for the corresponding s value,
4) Construct the network [Y] matrix in *numerical form*,
5) Compute $z=|\det(Y)|$ or $z=|\lambda_{min}(Y)|$ at that s value,
6) Plot z as a function of s. The poles can be located as the points of z=0. The poles' (i.e. modal) frequencies are very likely at interharmonic frequencies. The frequencies of phasor oscillation should be estimated using Eq. (3.3).

Fig. 23 shows the 2D frequency scan of a simple network consisting of an inductor L in series with a resistor R, which is then paralleled with a capacitor C. If R>0, the system is stable. Fig.23a shows the lowest point of the surface (i.e. the pole) is at the left-hand side. If R<0, the system is unstable, and the point of Y(s)=0 is at the right-hand side (Fig. 23b).

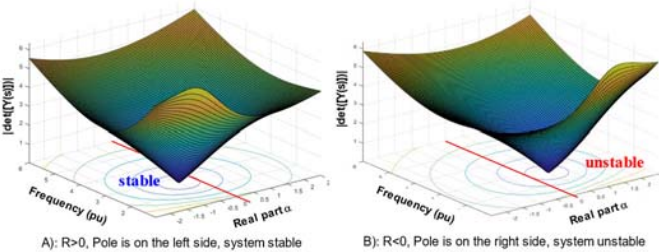

A): R>0, Pole is on the left side, system stable   B): R<0, Pole is on the right side, system unstable

Fig. 23: Sample results of a 2D frequency scan.

B. *Generator Participation Factor*

Once the poles are located, a numerical [Y] matrix at each unstable pole can be calculated by letting $s=s_{pole}=\alpha_{pole}+j2\pi f_{pole}$. The matrix can then be decomposed as follows:

$$[Y(s_{pole})] = [L][\Lambda][T] \quad (4.3)$$

where $[\Lambda]$ is the eigenvalue matrix, $[L]$ and $[T]$ are the left and right eigenvector matrices, and $[L]=[T]^{-1}$. Denoting the first eigenvalue $\lambda_1$ as the one approaching zero, [13,11] have shown that its corresponding right eigenvector, the 1st row of [T], can rank the buses/nodes where $f_{pole}$ oscillation can be excited and observed most easily. For example, if $|T_{15}|$ is the largest among $|T_{1i}|$ (i=bus#), bus 5 is the top location of excitability and observability for oscillations of that mode. This reasoning has led to the concept of bus participation factor [13,11]. It is defined as $P_{bus-i}=|L_{i1}|^2=|T_{1i}|^2$. This index has some useful applications such as optimal location for oscillation monitors. However, it is about buses and their easiness to excite/observe oscillations. Here we are interested in which generators cause oscillations. To this end, Eq.(4.1) & (4.3) are further processed in view that $\lambda_1^{-1}$ is much larger than other $\lambda^{-1}$s:

$$[\Delta V] = [Y]^{-1}[\Delta I] = [L][\Lambda]^{-1}[T][\Delta I]$$

$$= [L]\begin{bmatrix} \lambda_1^{-1} & 0 & 0 & 0 \\ 0 & \lambda_2^{-1} & 0 & 0 \\ 0 & 0 & ... & 0 \\ 0 & 0 & 0 & \lambda_n^{-1} \end{bmatrix}[T][\Delta I] \quad (4.4)$$

$$\approx [L]\begin{bmatrix} \lambda_1^{-1} & 0 & 0 & 0 \\ 0 & 0 & 0 & 0 \\ 0 & 0 & ... & 0 \\ 0 & 0 & 0 & 0 \end{bmatrix}[T][\Delta I] = \begin{bmatrix} L_{11} \\ L_{21} \\ ... \\ L_{n1} \end{bmatrix} U_1$$

where $\quad U_1 = \lambda_1^{-1} J_1, \quad J_1 = [T_{12} \; T_{13} \; ... \; T_{1n}][\Delta I]$

are called the modal voltage and modal current of $\lambda_1$ respectively. They can be understood as follows: perturbation $[\Delta I]$ can excite various modes of oscillations. A projection of the perturbation in the direction of mode 1 eigenvector, i.e. $J_1$, represents the perturbation pattern that only excites mode 1 oscillation. This modal current in turn will produce the mode 1 voltage through the modal admittance $\lambda_1$ as $U_1=\lambda_1^{-1}J_1$ [11].

According to Eq.(4.4), voltage at bus $i$ can be determined as $\Delta V_i = L_{i1}U_1$. Thus, if there is a generator connected at the bus, the generator will <u>consume</u> power at frequency $f_{pole}$ and this (interharmonic) power can be estimated as follows:

$$P_i = \Re\{\frac{|\Delta V_i|^2}{Z_{Gen-i}(s_{pole})}\} = \Re\{\frac{|L_{i1}|^2|U_1|^2}{Z_{Gen-i}(s_{pole})}\} = [G_{gen-i}(s_{pole})|L_{i1}|^2]|U_1|^2$$

where $Z_{gen}(s_{pole})$ is the generator's impedance at $s=s_{pole}$ (see Fig. 7a) and $G_{gen}$ is the conductance. Since $U_1$ is common to all buses, the following index can quantify the relative degree of interharmonic power <u>generation</u> of various generators:

$$P_{gen-i} = -G_{gen-i}(s_{pole})|L_{i1}|^2 \quad (4.5)$$

The above index is called "generator participation factor" for $s_{pole}$. A '–' sign is added so that positive $P_{gen-i}$ means interharmonic power is *generated*. Clearly, generators with negative $G_{gen-i}$ will generate interharmonic power ($P_{gen-i}>0$). A larger $P_{gen-i}$ means injection of more interharmonic power into the grid. As such, $P_{gen-i}$ can rank the degree of contribution or participation of a generator to $s_{pole}$ mode. If $P_{gen-i}<0$, it means the generator provides damping. An immediate application of this index is to find the most impactful IBRs, with their grid level interactions fully included, for controller tuning.

### C. Resonance Participation Factor

In addition to locating sources causing oscillations, stability investigators are also interested in finding network components that amplify oscillations. Oscillations are amplified through the mechanism of resonance. The mechanism also establishes the modal frequency $f_{pole}$. It is well known that resonance is caused by the energy exchanges among energy storge elements, i.e. capacitors and inductors. Since the amount of energy stored in or released from a resonating component is proportional to the reactive power "consumed" or "produced" by the component at the resonant frequency, it follows that the interharmonic reactive power can be used to quantify a component's participation in the $f_{pole}$ resonance. This index is called "Resonance Participation Factor", $P_{res}$, and it is derived as follows.

For a shunt component that has an impedance $Z_{shunt}(s_{pole})$ and is connected to bus $i$, the reactive power consumption of the component at mode $f_{pole}$ is

$$Q_i = \Im\{\frac{|\Delta V_i|^2}{Z_{Shunt}(s_{pole})}\} = [B_{shunt}(s_{pole})|L_{i1}|^2]|U_1|^2$$

where $B_{shunt}(s_{pole})$ is the susceptance of the component at $s_{pole}$. Its resonance participation factor is, therefore, defined as

$$P_{res-shunt@i} = B_{shunt}(s_{pole})|L_{i1}|^2$$

For a series component that has an impedance $Z_{series}(s_{pole})$ and is connected between buses $i$ and $j$, the reactive power consumption of the component at mode $\lambda_1$ is

$$Q_{ij} = \Im\{\frac{|\Delta V_i - \Delta V_j|^2}{Z_{Series}(s_{pole})}\} = [B_{series}(s_{pole})|L_{i1} - L_{j1}|^2]|U_1|^2$$

Thus, its resonance participation factor is

$$P_{res-series@ij} = B_{series}(s_{pole})|L_{i1} - L_{j1}|^2$$

If a component consists of both shunt and series elements, its resonance participation factor is the sum of those associated with each component. A large $|P_{res}|$ value means a strong participation in resonance. Components with positive and negative $P_{res}$ form a pair of resonant groups whose interaction results in the modal frequency $f_{pole}$.

An established method for identifying components critical to harmonic resonance is the eigen-sensitivity method [29,30]. This method could be extended to solve the current problem. It turns out that $P_{res}$ is proportional to the normalized eigen-sensitivity proposed in [30] (for reactive components). This outcome lends further support to the concept of $P_{res}$. The advantages of $P_{res}$ index over eigen-sensitivities are 1) it introduces a physical meaning, interharmonic reactive power, for resonant component identification, 2) $P_{res}$ can cover a component that has both series and shunt elements such as a transmission line, 3) it overcomes the requirement of $\lambda_1\neq0$ for the eigen-sensitivity calculation [30]. $\lambda_1\neq0$ is the case in harmonic resonance analysis, but this is not true for stability analysis per Eq.(4.2), and 4) $P_{res}$ is easier to calculate than the sensitivities.

The concept of using interharmonic reactive power for resonant components identification can also be added to the measurement-based oscillation monitoring and source location systems described in Section IV.A.

### D. An Illustrative Case Study

A case shown in Fig. 24 is used to demonstrate the model-base method. There are three identical induction generators (IGs) and they are connected to the grid with different line impedances. The system is analyzed using the proposed s-domain based method and the time-domain based simulation. Both studies show that there is an interharmonic around 49Hz and it causes 11Hz phasor oscillation. The min($\lambda$) surface of the 2D scan is shown in Fig. 25 and the lowest point is at about 49Hz. Table I shows the generator ranking based on normalized generator participation factors. It agrees with technical intuition since IG1 is the closest to the series capacitor. The ranking is in good agreement with the interharmonic power results obtained from time-domain simulation. The difference is caused mainly by signal processing issues since the simulated waveforms contain all spectral components. On the other hand, the generator participation factor includes only the impact of pure $f_{pole}$ component.

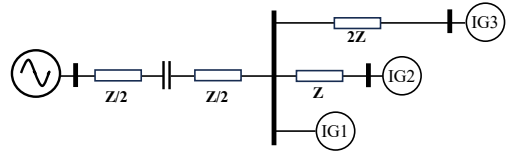

Fig. 24: A model-based test system.

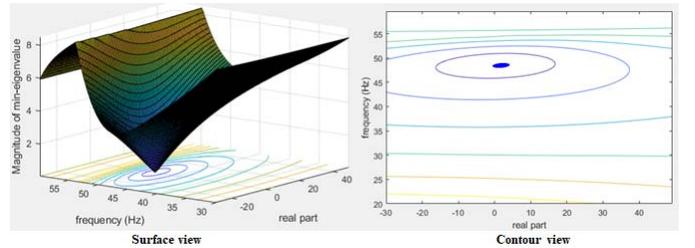

Fig. 25: Results of 2D frequency scan

Table I: Ranking of generator participations to 11Hz oscillation.

| Ranking indices | IG1 | IG2 | IG3 |
|---|---|---|---|
| Normalized gen. participation factor | 1.000 | 0.307 | 0.146 |
| Normalized inter. harmonic power | 1.000 | 0.341 | 0.153 |

### VI. CONCLUSIONS

Through analytical studies, mathematical derivations and field measurements, this paper has revealed that the generation and propagation of interharmonic powers is the primary driver of power system oscillations. Phasor oscillation is an appearance caused by interharmonics' modulating effect. The increasing oscillation frequencies caused by IBR integration necessitates to approach oscillation phenomena from the interharmonic perspective. Based on these understandings, two intuitively reasonable and yet mathematically rigorous interharmonic power-based methods for oscillation source location have been developed. The measurement-based method leads to a simple tool for oscillating source location and monitoring. By offering two key pieces of participation information, the model-based method significantly expands the capability of impedance-based stability assessment tools.

In spite of the above progress, additional research is still needed to achieve serious industry applications. For example, there is a need to research signal processing issues involving non-stationary waveforms and small interharmonic powers for the measurement-based method, such as window length, spectral leakage, and interharmonic grouping etc. For the model-based method, there is a need to establish the scan range. It is possible to improve the efficiency of 2D frequency scan by customizing the well-known power-iteration method for minimum eigenvalue determination.

The concept of decomposing oscillating phasors into interharmonics and corresponding rotating magnetic fields has some similarities to the concept of decomposing unbalanced phasors into sequence components and their rotating fields. The sequence decomposition has led to many innovations such as negative sequence protection and grounding transformers etc. It is possible that interharmonics approach to oscillations could lead to developments beyond oscillation source location, such as interharmonic relays and filters for oscillation mitigation and stability control. The response of PLL circuits to interharmonics may bring new insights on inverter instability. The extent of oscillation propagation in a grid can also be quantified using the eigenvector results.

## VII. APPENDIX A: DYNAMIC PHASOR VERSUS INTERHARMONICS

Dynamic phasors are a concept introduced to model varying voltage and current phasors. It has gained attractions in analyzing the behaviors of power electronic devices. There are also efforts in applying it to power system EMT simulations [31,32]. Oscillating phasor is a form of dynamic phasor. It is, therefore, useful to clarify which representation, dynamic phasor or interharmonic, are more suitable to investigate system oscillating behaviors. This Appendix uses two examples to investigate this issue.

### A. Calculating Current Caused by a Dynamic Voltage Source

This simple example is to calculate current I for the circuit shown in Fig. 26a. The voltage source E operates at 60Hz but has an oscillating magnitude at a frequency of $f_{os}=20Hz$. E can be represented as a dynamic phasor of

$$\vec{E}(t) = [1+0.1\cos(2\pi f_{os}t)]\angle 0^o \quad pu$$

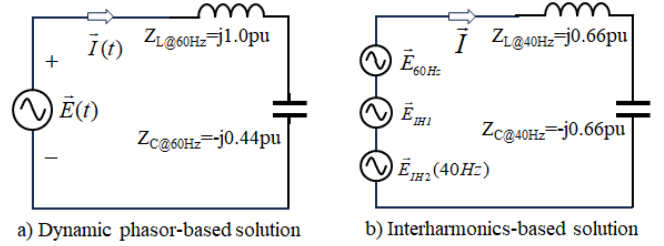

Fig. 26: A simple circuit to test dynamic phasor solution.

Based on traditional circuit method, the current can be solved as follows, which yields a dynamic current phasor.

$$\vec{I}(t) = \vec{E}(t)/(j1.0-j0.44) = [1.78+0.18\cos(2\pi f_{os}t)]\angle -90^o \quad pu$$

Note that the 60Hz impedances $Z_L$ and $Z_C$ are used, which is based on the understanding that source E is a 60Hz sinusoidal excitation. However, since E's magnitude oscillates, the frequency seen by L and C are not exactly 60Hz. If the oscillating frequency $f_{os}$ is relatively high, using 60Hz impedance values becomes questionable. To further investigate this issue, the true spectral components of E are determined, and the result is:

$$e(t) = \sqrt{2}[1+0.1\cos(2\pi f_{os}t)]\cos(2\pi f_1 t + 0)$$
$$= 1.414\cos(2\pi f_1 t) + 0.071\cos(2\pi f_{IH1}t) + 0.071\cos(2\pi f_{IH2}t)$$

where $f_{IH1}=80Hz$ and $f_{IH2}=40Hz$. The above shows that E is composed of three sources with frequencies of 60Hz, 80Hz and 40Hz respectively. All these sources have constant magnitudes so they can be represented accurately using standard (static) phasors. The current should be solved using the principle of superposition, i.e. it is equal to the time-domain summation of the currents produced by the three sources individually (Fig. 26b). Of particular interest is the current produced by the 40Hz source $E_{IH2}$. The current can be calculated as

$$\vec{I}_{IH2} = \vec{E}_{IH2}/(j1.0h-j0.44/h) = \vec{E}_{IH2}/(j0.66-j0.66) = \infty$$

where $h=f_{IH2}/f_1=40/60$ is a ratio to convert the $f_1$ impedance values into those at $f_{IH2}$. It can be seen that $E_{IH2}$ encounters a resonance (Fig. 26b). The resulting current is infinite. This is a phenomenon not predicted using the dynamic phasor method. In other words, if the oscillating frequency of a phasor is relatively high (which is true for IBRs), dynamic phasors have challenges to capture system behaviors.

### B. Measuring a Dynamic Phasor Voltage Using PMU

This example is to use the standard PMU algorithm to "measure" voltage E of Fig. 26a. It is to check if a PMU's output which is a dynamic phasor can reproduce E's behavior, and thus can be used for source location. The case of $f_{os}=42Hz$ is shown in Fig. 27. It can be seen that the PMU phasor oscillates at about 18Hz and it is clearly different from the behavior of E. This mismatch can be explained by the Nyquist theorem: the PMU phasor can only capture oscillations less than $30Hz(=f_1/2)$. The result shown in the figure is caused by the effect of aliasing since E osculates at 42Hz.

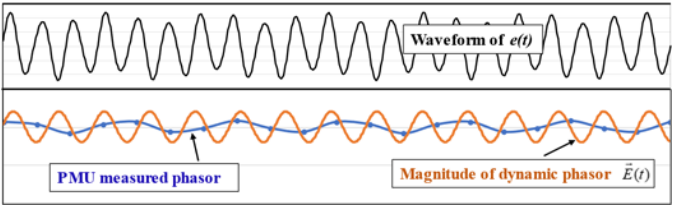

Fig. 27: Waveform e(t), original dynamic phasor, and "measured" phasor.

In summary, the above two examples have shown that dynamic fundamental frequency phasors have limitations to capture true oscillating behaviors. In fact, these limitations have been recognized by dynamic phasor researchers. Their proposed solution is to include dynamic phasors at higher frequencies, i.e. dynamic harmonic phasors [31,32]. According to their work, the window length used to define a dynamic harmonic phasor should be larger than $T_1(=1/f_1)$. The resulting dynamic "harmonic" phasors are actually interharmonic phasors. This development confirms that interharmonics are indeed needed to analyze oscillation behaviors.

## VIII. APPENDIX B: NECESSARY CONDITION OF PHASOR OSCILLATION

Let's consider the most general form of oscillating phasor:

$$\vec{V}_{phasor} = V(t)\angle\delta(t) \implies v(t) = \sqrt{2}V(t)\cos[\omega_1 t + \delta(t)]$$

i.e. both the magnitude and angle of the phasor can change. The waveform of this phasor can be expressed as:

$$\begin{aligned}v(t) &= \sqrt{2}V(t)\cos[\omega_1 t + \delta(t)] \\ &= \sqrt{2}V(t)\cos(\omega_1 t)\cos[\delta(t)] - \sqrt{2}V(t)\sin(\omega_1 t)\sin[\delta(t)] \\ &= m_a(t)\cos(\omega_1 t) + m_b(t)\sin(\omega_1 t)\end{aligned}$$

where $m_a(t) = \sqrt{2}V(t)\cos[\delta(t)]$, $m_b(t) = -\sqrt{2}V(t)\sin[\delta(t)]$.

Fourier transform is to be performed on $v(t)$. As explained earlier, the length of the data window $T_w$ for the transform shall cover one period of phasor oscillation, i.e. $T_w=1/f_{os}$. Let's further assume that this window includes k fundamental frequency cycles of data, i.e. $T_w=kT_1$, the corresponding frequency resolution of the transform is $f_w=1/T_w=f_1/k$. After performing Fourier transform on $m_a(t)$ and $m_b(t)$ using this window, both terms can be expressed as:

$$m_a(t) = \sum_{h=0}^{N} M_{ah}\cos(h\omega_w t + \alpha_h) \quad m_b(t) = \sum_{h=0}^{N} M_{bh}\cos(h\omega_w t + \beta_h)$$

Therefore,
$$\begin{aligned}v(t) &= \cos(\omega_1 t)\sum_{h=0}^{N} M_{ah}\cos(h\omega_w t + \alpha_h) \\ &+ \sin(\omega_1 t)\sum_{h=0}^{N} M_{bh}\cos(h\omega_w t + \beta_h)\end{aligned}$$

Accordingly, the Fourier component of $v(t)$ at frequency $nf_w$ (n=0,1,2…), $V_F$, can be calculated as follows:

$$V_F(nf_w) = \frac{2}{T_w}\int_0^{T_w} v(t)e^{-jn\omega_w t}dt$$

$$= \frac{2}{T_w}\int_0^{T_w} m_a(t)\cos(\omega_1 t)e^{-jn\omega_w t}dt + \frac{2}{T_w}\int_0^{T_w} m_b(t)\sin(\omega_1 t)e^{-jn\omega_w t}dt$$

If $m_a(t)$ and $m_b(t)$ are constant and equal to $M_a$ and $M_b$ respectively, i.e. there is no phasor oscillation, the following integral is always zero for different $n$,

$$\frac{2}{T_w}\int_0^{T_w} m_a(t)\cos(\omega_1 t)e^{-jn\omega_w t}dt = \frac{2}{T_w}M_a\int_0^{T_w}\cos(\omega_1 t)e^{-jn\omega_w t}dt = 0$$

In other words, if $m_a(t)$ is not a constant, the above term will be non-zero for at least some values of $n$. The same conclusion can be said to the $m_b(t)$ term. The implication is that some of the Fourier components $V_F(nf_w)$ are non-zero if the phasor oscillates. The frequencies of such components are:

$$nf_w = n/T_w = n/(kT_1) = (n/k)f_1, \quad n=1,2...$$

For example, if $k=60$ (i.e. 1 sec. window), the frequencies of $V_F$ are, $(1/60)f_1$, $(2/60)f_1$, …. Clearly these are interharmonic frequencies. $V_F$, therefore, represents the interharmonic components. In other words, an oscillating phasor always implies the existence of interharmonic components.

## IX. REFERENCES

[1]   F. P. de Mello, P. J. Nolan, T. F. Laskowski and J. M. Undrill, "Coordinated application of stabilizers in multimachine power systems", *IEEE Trans. on PAS*, vol. 99, no.8, pp. 892 – 901, Aug. 1980.
[2]   I. J. Perez-Arriaga, G. C. Verghese, and F. C. Schweppe, "Selective modal analysis with applications to electric power systems", Part I&II, *IEEE Trans. on PAS*, vol. 101, no. 9, pp. 3117-3134, Sep. 1982.
[3]   Y. Cheng, L. Fan, et al., "Real-world subsynchronous oscillation events in power grids with high penetrations of inverter-based resources," *IEEE Trans. on Power Systems,* vol. 38, no. 1, pp. 316-330, Jan. 2023. Analysis of asymmetrical faults in power systems using dynamic phasors
[4]   PES-TR110, "Forced Oscillations in Power Systems", Prepared by Oscillation Source Location Task Force, IEEE PES, May 2023.
[5]   B. Wang and K. Sun, "Location methods of oscillation sources in power systems: A survey," J. Modern Power Syst. Clean Energy, vol. 5, no. 2, pp. 151–159, Mar. 2017.
[6]   L. Chen, Y. Min, Y. Chen, and W. Hu, "Evaluation of generator damping using oscillation energy dissipation and the connection with modal analysis", *IEEE Trans. Power Systems*, vol.29, no.3, pp.1393-1402, May 2014.
[7]   Yuan Zhi, and V. Venkatasubramanian, "Analysis of energy flow method for oscillation source location", *IEEE Trans. on Power Systems,* vol.36, no.2, pp.1338-1349, 2021.
[8]   S. C. Chevalier, P. Vorobev, and K. Turitsyn, "Using effective generator impedance for forced oscillation source location," *IEEE Trans. on Power Systems*, vol. 33, no. 6, pp. 6264–6277, 2018.
[9]   X. Wang, L. Harnefors, and F. Blaabjerg, "Unified impedance model of grid-connected voltage-source converters" *IEEE Trans. on Power Electronics*, vol. 33, no. 2, pp. 1775–1787, 2017.
[10] L. Fan and Z. Miao, "Admittance-Based Stability Analysis: Bode Plots, Nyquist Diagrams or Eigenvalue Analysis?" in *IEEE Trans. on Power Systems*, vol. 35, no. 4, pp. 3312-3315, July 2020.
[11] W. Xu, Z. Huang, and Y. Cui, "Harmonic resonance mode analysis", *IEEE Trans. Power Delivery*, vol. 20, no. 2, pp. 1182–1190, Apr. 2005.
[12] E. Ebrahimzadeh, F. Blaabjerg, X. Wang, C.L. Bak, "Harmonic stability and resonance analysis in large PMSG-based wind power plants", *IEEE Trans. on Sustainable Energy*, vol.9, no.1, pp.12-23, Jan. 2018.
[13] Y. Zhan, X. Xie, H. Liu, H. Liu and Y. Li, "Frequency-domain modal analysis of the oscillatory stability of power systems with high-penetration renewables", *IEEE Trans. Sustainable Energy*, vol.10, no.3,

pp.1534-1543, July 2019.

[14] A. I. Semlyen, "s-domain methodology for assessing the small signal stability of complex systems in nonsinusoidal steady state," *IEEE Trans. on Power Systems*, vol. 14, no. 1, pp. 132–137, Feb 1999.

[15] IEC 61000-4-30: "Electromagnetic compatibility (EMC) – Part 4-30: Testing and measurement techniques – Power quality measurement methods", 2015.

[16] H.W. Dommel, "*EMTP Theory Book*", Microtran Power System Analysis Corporation, 2$^{nd}$ Edition, April 1996.

[17] IEEE Std C37.118.1-2011, "IEEE standard for synchrophasor measurements for power systems", Dec. 2011,

[18] J. Yong, T. Tayjasanant, W. Xu and C. Sun, "Characterizing voltage fluctuations caused by a pair of interharmonics," in *IEEE Trans. on Power Delivery*, vol. 23, no. 1, pp. 319-327, Jan. 2008,

[19] H.J. Marquez, "*Nonlinear Control Systems: Analysis and Design*", Published by John Wiley & Sons, Inc., Hoboken, New Jersy, 2003.

[20] B. Wen, D. Boroyevich, R. Burgos, P. Mattavelli, and Z. Shen, "Analysis of D-Q small-signal impedance of grid-tied inverters", *IEEE Trans. Power Electronics*, vol. 31, no. 1, pp. 675–687, Jan. 2016.

[21] W. Xu, H. Huang, X. Xie and C. Li "Synchronized Waveforms – a frontier of data-based power system and apparatus monitoring, protection, and control", *IEEE Trans. on Power Delivery*, vol. 37, no. 1, pp. 3-17, Feb. 2022.

[22] C. Li, W. Xu, and T. Tayjasanant, "Interharmonics: basic concepts and techniques for their detection and measurement," *Electric Power Systems Research*, vol. 66, No. 1, Pages 39-48, Jan. 2003.

[23] C. Li, "Unstable operation of photovoltaic inverter from field experiences", *IEEE Trans. Power Delivery*, vol. 33, no.2, pp. 1013-1015, Apr. 2018.

[24] R. Yacamini, "Power system harmonics. IV. Interharmonics," *Power Engineering Journal*, vol. 10, no. 4, pp. 185-193, Aug. 1996.

[25] IEEE Std 1453-2015, "IEEE recommended practice for the analysis of fluctuating installations on power systems," Oct. 2015.

[26] D. Zhang, W. Xu and A. Nassif, "Flicker source identification by interharmonic power direction," *Canadian Conference on Electrical and Computer Engineering, 2005.*, Saskatoon, 2005, pp. 549-552.

[27] P. Axelberg, M. Bollen, I. Gu, "Trace of flicker sources by using the quantity of flicker power", *IEEE Trans. Power Delivery*, vol.23, no.1, pp. 465-471, Jan. 2008.

[28] A. B. Nassif, J. Yong, H. Mazin, X. Wang and W. Xu, "An impedance-based approach for identifying interharmonic sources," in *IEEE Trans. on Power Delivery*, vol. 26, no. 1, pp. 333-340, Jan. 2011.

[29] B. Porter and R. Crossley, Modal Control Theory and Applications, London, pp. 21-43, 1972.

[30] Z. Huang, Y. Cui and W. Xu, "Application of Modal Sensitivity for Power System Harmonic Resonance Analysis," in *IEEE Transactions on Power Systems*, vol. 22, no. 1, pp. 222-231, Feb. 2007.

[31] A. M. Stankovic and T. Aydin, "Analysis of asymmetrical faults in power systems using dynamic phasors," in *IEEE Trans. on Power Systems*, vol. 15, no. 3, pp. 1062-1068, Aug. 2000.

[32] J. J. Rico, M. Madrigal and E. Acha, "Dynamic harmonic evolution using the extended harmonic domain," in *IEEE Trans. on Power Delivery*, vol. 18, no. 2, pp. 587-594, April 2003.